\documentclass[prd,twocolumn,aps,showpacs,nofootinbib,nobibnotes,superscriptaddress,preprintnumbers]{revtex4-1}
\usepackage{epsfig}
\usepackage{graphicx}
\usepackage{bm}
\usepackage[usenames, dvipsnames]{color}
\usepackage{dcolumn}   
\usepackage[spanish,english]{babel}
\usepackage{bm}     
\usepackage{bbm}       
\usepackage{amssymb}  
\usepackage{amsmath}
\usepackage{latexsym}
\usepackage{float}
\usepackage{ifthen}
\usepackage{caption,subfig}
\usepackage{enumerate}
\usepackage{url}
\usepackage{caption,subfig}
\usepackage{amsopn}
\usepackage[colorlinks,citecolor=blue,linkcolor=blue,urlcolor=blue]{hyperref}

\usepackage{amsfonts}
\usepackage{multirow}
\usepackage{array}
\usepackage{booktabs}
\usepackage{rotating}

\newcommand{\D}{\mathrm{d}}

\usepackage{ulem}
\def\clap#1{\hbox to 0pt{\hss#1\hss}}

\def\bea{\begin{eqnarray}}
\def\eea{\end{eqnarray}}
\def\be{\begin{equation}}
\def\ee{\end{equation}}
\def\mpl{M_{\rm P}}

\newcommand{\dd}{{\rm d}}

\newcommand{\QQ}{\mathbb{Q}}
\newcommand{\QGR}{\mathring{\mathbb{Q}}}

\begin{document}

\title{Non-linear extension of non-metricity scalar for MOND}

\author{Fabio D'Ambrosio} \email{fabioda@phys.ethz.ch}
\author{Mudit Garg} \email{gargm@student.ethz.ch}
\author{Lavinia Heisenberg} \email{lavinia.heisenberg@phys.ethz.ch}
\affiliation{Institute for Theoretical Physics,
ETH Zurich, Wolfgang-Pauli-Strasse 27, 8093, Zurich, Switzerland}

\date{\today}

\begin{abstract}
General Relativity enjoys the freedom of different geometrical interpretations in terms of curvature, torsion or non-metricity. Within this geometrical trinity, a simpler geometrical formulation of General Relativity manifests itself in the latter, where gravity is entirely attributed to non-metricity. In this Letter, we consider non-linear extensions of Coincident General Relativity $f(\QGR)$ for phenomenological applications on both cosmological as well as galactic scales. The theory not only delivers dark energy on large scales but also recovers MOND on galactic scales, together with implications for the early universe cosmology. To the best of our knowledge, this represents the first relativistic, covariant, and ghost-free hybrid-formulation of MOND which recovers both, General Relativity and MOND in the appropriate limits and reconciles expected cosmological behavior. We further illustrate that previous bimetric formulations of MOND generically suffer from ghost instabilities and $f(\QGR)$ crystalizes as a unique ghost-free theory.
\end{abstract}


\maketitle


\section{Introduction}
Since its inception, General Relativity is perceived as the geometrical property of spacetime. The origin of this interpretation comes from the equivalence principle. Einstein has chosen to describe gravity in terms of the curvature of spacetime. On an equal footing he could have embraced the much richer structure of the affine sector. Even if it is not widely known, General Relativity admits two alternative and fully equivalent representations in flat geometries, pushing forward a geometrical trinity of gravity \cite{BeltranJimenez:2019tjy} (see also \cite{Heisenberg:2018vsk}). Apart from its traditional representation in terms of curvature, General Relativity can be attributed to torsion or identified with non-metricity. The former represents the Teleparallel Equivalent of GR (TEGR) in Weitzenb\"ock spaces \cite{Aldrovandi:2013wha} whereas flat and torsion-free spacetimes can accommodate a Symmetric Teleparallel Equivalent of GR (STEGR) \cite{BeltranJimenez:2017tkd}.

Independently of its formulation, General Relativity requires the introduction of a dark matter component in the universe in order to describe, among other things, the right phenomenology on galactic scales. This component is believed to consist of weakly interacting cold matter that does not directly couple to light. Its gravitational manifestation is, however, quite pertinent, making up almost 30\% of the energy budget of the universe. Its presence enables us to explain for instance the flat rotation curves of spiral galaxies, which according to Newtonian physics should have substantially decreased \cite{Rubin:1982kyu}. A widely held belief is that the disks of galaxies are embedded in quasi-spherical potentials of huge halos of dark matter. Even though not completely indisputable, some concerns were raised that some of the predictions of the cold dark matter model are not observed, like the generic tendency of dark matter cusps in central regions of galaxies and the phase-space correlation of galaxy satellites. Similarly, it is claimed that the standard cold dark matter model fails to explain some of the relevant galactic observations, like the baryonic Tully-Fisher relation for spiral galaxies and the Faber-Jackson relation for elliptical galaxies as well as the tight correlation between the luminous versus dynamical mass. 

In an attempt to account for dark matter phenomenology without introducing additional matter components, the MOdified Newtonian Dynamics (MOND) model was proposed \cite{Milgrom:1983ca}. The trade-off is that it introduces modified laws of motion for small accelerations. Even though some of the challenges of cold dark matter at galactic scales are addressed, MOND faces tenacious difficulties in explaining the dark matter distribution at large scales of galaxy clusters \cite{Pointecouteau:2005mr} and its rigidity in accounting for the phenomenology in the solar system \cite{Blanchet:2010it}. It has been a challenging task to construct a consistent relativistic hybrid theory that reconciles the MOND phenomenology at galactic scales with the $\Lambda$CDM phenomenology on large, cosmological scales. Lorentz breaking attempts include Einstein-Aether theories \cite{Zlosnik:2006zu} and TeVeS \cite{Bekenstein:2004ne}. Another approach is the bimetric formulation of MOND \cite{Milgrom:2009gv}. Instead of replacing dark matter, it was also contemplated to consider a new form of dark matter which mimics MOND on galactic scales -- the so-called dipolar dark matter \cite{Blanchet:2006yt}. It relies on a mechanism of gravitational polarization in the presence of two different dark matter species coupled to different gravitational potentials. Therefore, a natural realization arises in bimetric formulations of gravity \cite{DipolarDM}, where the two species of dark matter components couple separately either to the $g$ metric or to the $f$ metric and are linked via an internal vector field.

In this Letter we consider a non-linear extension of non-metricity scalar $f(\QGR)$ within the geometrical formulation of gravity as a relativistic, covariant, Lorentz invariant, and ghost-free theory that recovers both, General Relativity and MOND in certain limits and admits $\Lambda$CDM as a cosmological evolution on large scales. We show explicitly the successful recovery of MOND in the non-relativistic limit of the theory and discuss a few exemplary ans\"atze relevant for self-accelerating solutions. Some of these ans\"atze have important relevance for both, the deep MOND regime and the early universe cosmology. In \cite{Milgrom:2019rtd} it was shown that the Bimetric Modification Of Newtonian Dynamics (BIMOND) \cite{Milgrom:2009gv} corresponds to $f(\QGR)$ theory in the coincident gauge if the second metric of bigravity is forced to be Minkowski. We confirm this finding and further systematically show that it is not possible to have a ghost-free BIMOND theory which also recovers the MOND phenomenology in the non-relativistic limit, except in the unique case where it becomes the $f(\QGR)$ theory. This would suggest to abandon the bimetric formulation of MOND and consider solely $f(\QGR)$ as a promising relativistic and covariant theory for MOND.

\section{$f(\QGR)$ theory}
The geometrical interpretation of gravity allows three distinctive but entirely equivalent formulations of General Relativity. We chose to concentrate on the representation based on non-metricity. It will be assumed that the connection and the metric are independent geometrical quantities on the manifold under consideration. The bedrock of this formulation is teleparallelism, which is characterized by the vanishing of the curvature tensor,
\begin{equation}
R^\alpha{}_{\beta\mu\nu}=2\partial_{[\mu}\Gamma^\alpha{}_{\nu]\beta}+2\Gamma^\alpha{}_{[\mu|\lambda|} \Gamma^\lambda{}_{\nu]\beta}\overset{!}{=}0.
\label{eq:zeroR}
\end{equation}
This in turn implies that the affine connection is expressible as
\begin{equation}
\Gamma^\alpha{}_{\mu\beta}=(\Lambda^{-1})^\alpha{}_\rho\partial_\mu\Lambda^\rho{}_\beta,
\label{eq:Telecon}
\end{equation}
where $\Lambda^\alpha{}_\beta\in GL(4,\mathbb{R})$. STEGR further requires the vanishing of the torsion tensor 
\begin{equation}
T^\alpha{}_{\mu\beta}=2\Gamma^\alpha{}_{[\mu\beta]}=2(\Lambda^{-1})^\alpha{}_\rho\partial_{[\mu}\Lambda^\rho{}_{\beta]}\overset{!}{=}0.
\label{eq:zeroT}
\end{equation}
Combining conditions \eqref{eq:zeroR} and \eqref{eq:zeroT} has far reaching consequences. The teleparallel connection becomes a pure diffeomorphism, 
\begin{equation}
\Lambda^\alpha{}_\beta=\partial_\beta\xi^\alpha,
\label{GammaSTEGR}
\end{equation}
where the arbitrary $\xi^\alpha$'s can be identified with St\"uckelberg fields restoring covariance. The special choice of identifying them with a coordinate transformation $\xi^\alpha=x^\alpha$ corresponds to the coincident gauge with $\Gamma^\alpha{}_{\mu\beta}=0$. The remaining geometrical object, the non-metricity, 
\begin{equation}
Q_{\alpha\mu\nu}=\nabla_\alpha g_{\mu\nu}=\partial_\alpha g_{\mu\nu}-2(\Lambda^{-1})^\lambda{}_\rho\partial_\alpha\Lambda^\rho{}_{(\mu} g_{\nu)\lambda}
\end{equation}
constitutes the fundamental building block of our theory. Since the non-metricity tensor contains three indices, one cannot construct a scalar quantity at linear order. Therefore, the construction starts at quadratic order with five independent contractions, which reflects the symmetric nature of the last two indices of the non-metricity tensor. Similarly, this sets up two independent traces $Q_\mu:=Q_{\mu\alpha}{}^\alpha$ and $\bar{Q}_\mu:=Q^{\alpha}{}_{\alpha\mu}$. The quadratic action reads
\begin{equation}
\mathcal{S}=-\frac12\mpl^2\int\dd^4x\sqrt{-g}\,\QQ,
\end{equation}
where the short hand notation $\QQ$ stands for
\begin{align}\label{eqSIIgen}	
\QQ&=
c_{1}Q_{\alpha\mu\nu}Q^{\mu\alpha\nu} +
	c_{2}Q_{\alpha\mu\nu}Q^{\mu\nu\alpha} +
	c_{3}Q_{\mu}Q^{\mu} \nonumber\\
	&+
	c_{4}\bar{Q}_{\mu}\bar{Q}^{\mu}+
	c_{5}Q_{\mu}\bar{Q}^{\mu}.
\end{align}
 The theory becomes equivalent to General Relativity for the parameter choices
 \begin{equation}
 c_1=\frac14, \,c_2=-\frac12, \,c_3=-\frac14, \,c_4=0, \,c_5=\frac12,
 \end{equation}
which we will denote by $\QGR$. The Einstein-Hilbert action is equivalent to the quadratic non-metricity action in terms of $\QGR$ which satisfies the duality relation
 \begin{equation}
 \mathcal{R}=-\QGR-\mathcal{D}_\alpha (Q^\alpha-\bar{Q}^\alpha),
  \end{equation}
  where $\mathcal R$ is the Ricci scalar of the Levi-Civita connection and $\mathcal{D}_\alpha$ stands for the metric-compatible covariant derivative.

In this Letter, we consider a non-linear extension of the non-metricity scalar in form of a general function,
\begin{equation}\label{fQtheory}
\mathcal{S}=-\frac12\mpl^2\int{\rm d}^4x\sqrt{-g}\,f(\QGR),
\end{equation}
supplemented by the standard matter action. Introducing the non-metricity conjugate
\begin{align}\label{NMcon}	
P^\alpha{}_{\mu\nu}=-\frac{Q^\alpha{}_{\mu\nu}}4+\frac{Q_{(\mu\nu)}{}^\alpha}2-\frac{\delta^\alpha_{(\mu}Q_{\nu)}}4
	+\frac{g_{\mu\nu}}4(Q^\alpha-\bar{Q}^\alpha),
\end{align}
the metric field equations can simply be expressed as
\begin{equation}
\frac{f}2\delta^\alpha_\beta+f'P^{\alpha\mu\nu}Q_{\beta\mu\nu}+\frac{2}{\sqrt{-g}}\nabla_\mu(\sqrt{-g}f'P^{\mu\alpha}{}_\beta)=\frac{T^\alpha{}_\beta}{\mpl^2}
\end{equation}
and similarly the connection field equations read
\begin{equation}
\nabla_\alpha\nabla_\beta(\sqrt{-g}f'P^{\alpha\beta}_{\mu})=0.
\end{equation}
The stress energy tensor of the standard matter fields satisfies $\mathcal{D}_\alpha T^\alpha{}_\beta=0$ and no matter couplings to the connection are considered, i.e., the hypermomentum vanishes.
The model \eqref{fQtheory} represents a promising extension with rich cosmological implications \cite{Jimenez:2019ovq}. Here, we consider it as a relativistic, ghost-free covariant formulation of MOND
that can mimic $\Lambda$CDM on cosmological scales and MOND on galactic scales. It can even have interesting implications for early universe cosmology.

\section{The non-relativistic limit}
In this section we are interested in the recovery of the MOND phenomenology on galactic scales. To that end, we study the non-relativistic limit of the $f(\QGR)$ theory. 
We follow \cite{Milgrom:2019rtd} and adapt it to our case. As a reminder, the teleparallelism and torsion-freeness conditions impose
\begin{align}\label{eqSIIgen}	
\Gamma^{\mu}{}_{\alpha\nu}&=\frac{\partial x^\mu}{\partial \xi^\lambda}\partial_\alpha\partial_\nu\xi^\lambda \nonumber\\
Q_{\alpha\mu\nu}&=\partial_\alpha g_{\mu\nu}-2\frac{\partial x^\rho}{\partial\xi^\lambda}\partial_\alpha\partial_{(\mu}\xi^\lambda g_{\nu)\lambda}.
\end{align}
In the coincident gauge, i.e. $\xi^\alpha=x^\alpha$, the connection vanishes and the non-metricity tensor simplifies to $Q_{\alpha\mu\nu}=\partial_\alpha g_{\mu\nu}$. The quadratic non-metricity scalar $\QGR$ in this gauge simply becomes
\begin{equation}
\QGR=g^{\mu\nu}\left( \{^\alpha_{\beta\mu}\}  \{^\beta_{\nu\alpha}\}-  \{^\alpha_{\beta\alpha}\}  \{^\beta_{\mu\nu}\}\right),
\end{equation}
where $\{^\alpha_{\beta\mu}\}$ are the Christoffel symbols of the metric $g_{\mu\nu}$.
We rewrite our action \eqref{fQtheory} as
\begin{equation}\label{fQtheoryAns}
\mathcal{S}=\frac12\mpl^2\int{\rm d}^4x\sqrt{-g}\,2a_0^2f\left(\frac{-\QGR}{2a_0^2}\right),
\end{equation}
where $a_0$ will later represent the MOND acceleration parameter on galactic scales. For the function $f$ we will assume the ansatz $f\left(\frac{-\QGR}{2a_0^2}\right)=\frac{-\QGR}{2a_0^2}+\mathcal{M}\left(\frac{-\QGR}{2a_0^2}\right)$. The non-relativistic limit captures the relevant physics of slowly moving sources in a weak field approximation and we can therefore perturb the metric over a flat background as
\begin{equation}
    g_{\mu\nu}=\eta_{\mu\nu}-2\phi\delta_{\mu\nu}+h_{\mu\nu},
\end{equation}
and assume the stress energy tensor of the source to be of the form $T_{00}=\rho$. Note that $2\phi=\eta_{00}-g_{00}$ and $h_{00}=0$. The field equations in this limit become
\begin{align}\label{eqoNRlimit}
    \partial_i\left(f'\left(\frac{-\QGR}{2a_0^2}\right)S^i_{~\mu\nu}\right)=8\pi G\rho\delta_{~\mu}^0\delta_{~\nu}^0,
\end{align}
where the non-metricity scalar in the non-relativistic limit reads
\begin{align}
\QGR&=-2(\vec{\nabla}\phi)^2-\frac{1}{4}[h^{ij,k}(h_{ij,k}-2h_{ik,j})-{h}^{,k}({h}_{,k}-2h^j_{k,j})]\nonumber\\
&+\frac{1}{4}(h_{0i,j}-h_{0j,i})(h^{0i,j}-h^{0j,i}),
\end{align}
and $S^i_{\mu\nu}$ are shorthand notations for
\begin{align}
   S^i_{~00}=&~2\phi^{,i}+\frac{1}{2}(h^{ij}_{~,j}-{h}^{,i}),\nonumber\\
  S^i_{~jk}=&\frac{1}{2}(h^{i}_{~j,k}+h^{i}_{~k,j}-h_{jk}^{~~,i})\nonumber\\
  &+\frac{1}{4}[2\delta_{jk}({h}^{,i}-h^{im}_{~,m})-\delta^{i}_{~j}{h}_{,k}-\delta^{i}_{~k}{h}_{,j}], \nonumber\\
  S^i_{~0j}=&\frac{1}{2}(h^{0i}_{~~,j}-h_{0j}^{~~,i}),
\end{align}
where $h=h^i_{~i}$. Assuming boundary conditions at infinity for which $h_{0i}\to0$ and $h_{ij}\to0$ fast enough, the specific $\{^._{..}\}^2$ dependence of the non-metricity scalar $f\left(\frac{-\QGR}{2a_0^2}\right)$ allows for solutions with vanishing $h_{0i}$ and $h_{ij}$ components.
Hence, the temporal components of the equations of motion \eqref{eqoNRlimit} result in the non-linear MOND Poisson equation
\begin{equation}
    \vec{\nabla}\left(f'(|\vec{\nabla}\phi|^2/a_0^2)\vec{\nabla}\phi\right)=4\pi G\rho.
\end{equation}
In terms of our functional ansatz this becomes
\begin{equation}
    \vec{\nabla}^2\phi+ \vec{\nabla}\left(\mathcal{M}'(|\vec{\nabla}\phi|^2/a_0^2)\vec{\nabla}\phi\right)=4\pi G\rho.
\end{equation}
The general function of the non-metricity scalar has to satisfy the following conditions:
\begin{itemize}
    \item[1.] For $a_0\to0$ General Relativity should be recovered. This implies that for $z:=-\frac{ \QGR}{2a_0^2}\to \infty$ the general function converges to $f(z)\to z+\mathcal{M}_\infty$, with $\mathcal{M}_\infty$ some dimensionless constant, or in other words $\mathcal{M}'(z) \to 0$. Hence, our initial action \eqref{fQtheoryAns} becomes
    \begin{equation}
    \mathcal{S}=\frac{1}{16\pi G}\int \D^4x\sqrt{-g}\left[- \QGR+2a_0^2\mathcal{M}_\infty\right],
    \end{equation}
    which is just Coincident General Relativity with a cosmological constant $2a_0^2\mathcal{M}_\infty$.\footnote{We would like to thank Mordehai Milgrom to pointing out that this reflects the $a_0$-cosmology coincidence where the cosmological constant naturally coincides with the MOND parameter already in the ghostly BIMOND formulations.} Similarly, its non-relativistic limit recovers the standard Newton-Poisson equation $\vec{\nabla}^2\phi=4\pi G\rho$. Hence, for $a_0\to0$ one recovers $\Lambda$CDM on large scales and Newtonian physics in the non-relativistic limit on small scales.
    \item[2.] In the deep MOND regime, i.e., $a_0\to\infty$, while keeping $Ga_0$ fixed, scale invariance demands that for $z\to 0$ we must have $f(z)\to c z^{3/2}+\mathcal{M}(0)$. Thus, for $a_0\to\infty$ one recovers the correct MOND phenomenology for specific forms of the general function $f$.
\end{itemize}

\section{Cosmological implications}
In the previous section we have seen that the promising $f(\QGR)$ theory successfully recovers both, the Newtonian and the MOND limit. Another distinctive and unique property of this theory is that it can play the role of a hybrid model and connect to the correct cosmological behavior as well. It has relevance on late-time as well as on early-time cosmology. For the general ansatz $f\left(\frac{-\QGR}{2a_0^2}\right)=\frac{-\QGR}{2a_0^2}+\mathcal{M}\left(\frac{-\QGR}{2a_0^2}\right)$ we have seen that General Relativity with a cosmological constant $2a_0^2\mathcal{M}_\infty$ is recovered for $a_0\to0$. In this limit we would have exactly the same cosmological behavior as in the $\Lambda$CDM model, where $2a_0^2\mathcal{M}_\infty$ plays the role of $\Lambda$.

For a homogeneous and isotropic ansatz $\D s_g^2=-N(t)^2\D t^2+a(t)^2\D x^2$ the cosmological background equations of motion become \cite{Jimenez:2019ovq}
\begin{eqnarray}\label{BGeomfQ}
6f'H^2-\frac12f&=&8\pi G\rho \label{eqFrid}, \nonumber\\
\big(12H^2f''+f'\big)\dot{H}&=&-4\pi G(\rho+p)\, ,
\end{eqnarray}
with $\QGR=6H^2/N^2$. Even after fixing the
coincident gauge, we are allowed to set N = 1 due to accidental symmetry
of the cosmological background in $f(\QGR)$ theory. For the ansatz of the function $\mathcal{M}$ of the form
\be
\mathcal{M}\left(\frac{-\QGR}{2a_0^2}\right)=-6\beta a_0^2\left(\frac{\QGR}{6a_0^2}\right)^\alpha
\ee
the Friedmann equation turns into
\be
H^2\left[1+(1-2\alpha)\beta\left(\frac{H^2}{a_0^2}\right)^{\alpha-1}\right]=\frac{8\pi G}{3}\rho\,.
 \ee
Models with $\alpha>1$ have important implications for the early universe with possible corrections to inflationary solutions. Since the deep MOND regime, characterized by $a_0\to\infty$, requires $f(z)\to c z^{3/2}+\mathcal{M}(0)$ for $z\to 0$, choosing $\alpha=3/2$ will not only successfully recover MOND but also have interesting consequences for the early universe. 

Summarizing we can say that
\begin{itemize}
    \item[a)] in the limit $a_0\to0$ General Relativity with a cosmological constant is recovered which therefore yields the correct $\Lambda$CDM phenomenology for the late-time universe and simultaneously guarantees the recovery of the Newtonian behavior in the non-relativistic limit;
  \item[b)] in the limit $a_0\to\infty$ the deep MOND regime is obtained, simultaneously with relevant implications for early universe cosmology.
\end{itemize}
To the best of our knowledge, this is one of the first theories which is consistently applicable to a wide range of phenomenological frontiers. From a theoretical point of view it is also a very appealing theory since it neither introduces ghostly degrees of freedom nor explicitly breaks Lorentz symmetry.

\section{BIMOND}
\subsection{The original formulation of BIMOND}
 In \cite{Milgrom:2019rtd} it was shown that constrained BIMOND \cite{Milgrom:2009gv} is equivalent to $f(\QGR)$ theory in the coincident gauge, provided the second metric of BIMOND is forced to be flat. We have explicitly checked and confirmed that the constrained version of BIMOND indeed recovers $f(\QGR)$ in the coincident gauge. However, all the other non-constrained versions of BIMOND suffer from the Boulware-Deser ghost and hence do not represent viable relativistic and covariant embeddings for MOND.
 
 BIMOND is a bimetric covariant relativistic formulation of  MOND which relies on two metrics, $g_{\mu\nu}$ and $f_{\mu\nu}$ \cite{Milgrom:2009gv}. The problematic feature of this theory is twofold: The two metrics interact via derivative couplings and their potential interactions do not belong to the ghost-free dRGT potentials \cite{deRham:2010kj}. As it is the case for bigravity \cite{Hassan:2011zd}, each metric carries its own kinetic term. However, in BIMOND the individual connections mix. Consider the following quantity 
 \begin{equation}
    C^\alpha_{\mu\nu} := \{^\alpha_{\beta\mu}\} -\langle^\alpha_{\beta\mu}\rangle
\end{equation}
which represents the difference between the Levi-Civita connection $\{^\alpha_{\beta\mu}\}$ of the $g_{\mu\nu}$ metric and the Levi-Civita connection $\langle^\alpha_{\beta\mu}\rangle$ of the $f_{\mu\nu}$ metric. At quadratic order one can construct the scalar quantity
\begin{equation}
    \Upsilon :=g^{\mu\nu}\left(C^\gamma_{\mu\lambda}C^\lambda_{\nu\gamma}-C^\gamma_{\mu\nu}C^\lambda_{\lambda\gamma}\right),
\end{equation}
with the help of which the suggested action in \cite{Milgrom:2019rtd} reads
\begin{align}\label{actBimond}
    \mathcal{S}&=\int \D^4x\Big[\mpl^2\sqrt{-g}\mathcal{R}_g+M_f^2\sqrt{-f}\mathcal{R}_f \nonumber\\
    &+2a_0^2(gf)^{\frac{1}{4}}P\left(\left(\frac{g}{f}\right)^{\frac14}\right)\mathcal{M}\left(-\frac{\Upsilon}{2a_0^2}\right)\Big].
\end{align}
A consistent bigravity theory should only contain seven propagating degrees of freedom, corresponding to two for the massless spin-2 field and five for the massive spin-2 field. The action \eqref{actBimond}, however, contains at least eight degrees of freedom, one of which is the Boulware-Deser ghost which renders the theory sick and neither viable for cosmological nor astrophysical applications. The two sources for the Boulware-Deser ghost in this action are
\begin{itemize}
    \item[1.] The term $(gf)^{\frac{1}{4}}P(X)$, with $X=(g/f)^{1/4}$, apart from being the standard volume element $\sqrt{-g}$ for $P(X)=X$ , is not part of the unique ghost-free dRGT potentials, therefore does not provide the extra constraint to remove the Boulware-Deser ghost.
    \item The no-go theorem presented in \cite{deRham:2013tfa} clearly states that kinetic interactions free from the Boulware-Deser ghost do not exist. The function $\mathcal{M}\left(-\frac{\Upsilon}{2a_0^2}\right)$, which is responsible for the connection mixing, introduces such kinetic couplings with the Boulware-Deser ghost.
 \end{itemize}
 Already perturbations over a flat background reveal that the term $(gf)^{\frac{1}{4}}P(\left(g/f\right)^{1/4})$ does not have the correct potential structure. Actually, it does not even account for the right structure for the linear Fierz-Pauli mass term. The linear mass terms have a detuning between $[h]^2$ and $[h^2]$, where $[\cdot]$ denotes the trace of the metric perturbations. Furthermore, the function $\mathcal{M}\left(-\Upsilon/(2a_0^2)\right)$ gives rise to kinetic mixing of the spin-2 perturbations for the Fierz-Pauli Lagrangian. Thus, the Boulware-Deser ghost is already excited at linear order and in terms of the helicity-0 mode it would scale as $(\Box\pi)^2/a_0^2$. Since the ghost is a very light degree of freedom one cannot use the model as an effective field theory. The MOND effects will unavoidably be accompanied by the ghost.
 
 Another quick way to see the manifestation of the Boulware-Deser ghost is through mini-superspace models where the two metrics take the form
 \begin{align}\label{miniSuperspace}
   \D s_g^2&=-N_g(t)^2 \D t^2+a_g(t)^2 \D x^2 \nonumber\\
 \D s_f^2&=-N_f(t)^2 \D t^2+a_f(t)^2 \D x^2.
\end{align}
The presence of $\Upsilon$ renders the lapse functions dynamical while the function $P(X)$, with $X=(g/f)^{1/4}$, multiplied by the prefactor $(gf)^{\frac{1}{4}}$ gives rise to a non-linear appearance of the lapse functions, except in the trivial case $P(X)=X$, where the product simply becomes $\sqrt{-g}$. We have explicitly checked that the resulting Hamiltonian is not linear in either one of the lapse functions and even carries dynamics for both of them in the case of kinetic mixing. Since there is no shift vector field in this special case of mini-superspace, the only way to generate a constraint that can remove the Boulware-Deser ghost is if the lapses appear linearly in the Hamiltonian. This is only achieved if $f_{\mu\nu}$ is forced to be $\eta_{\mu\nu}$ and the potential coupling to be $P(X)=X$. We also performed a detailed cosmological perturbation analysis, where we perturbed the metrics $g_{\mu\nu}=\bar{g}_{\mu\nu}+h_{\mu\nu}$ and $f_{\mu\nu}=\bar{f}_{\mu\nu}+w_{\mu\nu}$ on top of the mini-superspace backgrounds \eqref{miniSuperspace}. Apart from the special case with $f_{\mu\nu}=\eta_{\mu\nu}$ and $P(X)=X$, quite generically, BIMOND interactions gave rise to the propagation of $\psi$ and $E$ of the spatial metric perturbations of the massive spin-2 sector, where one of them represents the Boulware-Deser ghost or gave rise to strong coupling problems for some specific functions.

\subsection{Constrained BIMOND: $f(\QGR)$}
In the previous subsection we have seen that the BIMOND model in its original formulation without any restrictions suffers from the Boulware-Deser ghost. The only way to avoid this pathological degree of freedom is the restriction $f_{\mu\nu}=\eta_{\mu\nu}$ and $P(X)=X$. In this case one exactly recovers the $f(\QGR)$ Lagrangian in the coincident gauge. Already in the mini-superspace model one observes that the dependence on the lapse function becomes linear after passing to the Hamiltonian formalism. To see this, let us rewrite our initial action $\mathcal{S}=\int{\rm d}^4x\sqrt{-g}\,f(\QGR)$ as
\begin{equation}
\mathcal{S}=\int{\rm d}^4x\sqrt{-g}\,\left[f(q) -\lambda(q-\QGR)\right],
\end{equation}
with $f'(q)=\lambda$. The action on the mini-superspace then becomes
\begin{equation}
\mathcal{S}=\int{\rm d}^4x\sqrt{-g}\,\left[f(q) -\lambda\left(q+\frac{6\dot{a}^2}{a^2N^2}\right)\right].
\end{equation}
As apparent from the action, the lapse first enters non-linearly. However, by computing the conjugate momenta
 \begin{align}\label{conjMom}
   \Pi_N&=0, \qquad  \Pi_\lambda=0\nonumber\\
   \Pi_q&=0, \qquad  \Pi_a=-\frac{12a\lambda}{N}\dot{a}
\end{align}
and using the momentum conjugate to the scale factor to express $\dot{a}$ in terms of $\Pi_a$, we can write the primary Hamiltonian as
 \begin{align}
H=\int{\rm d}^3x\,\Big\{L_N \Pi_N+L_\lambda \Pi_\lambda+L_q \Pi_q \nonumber\\
-a^3N\left(f(q)-\lambda\left(q-\frac{\Pi_a^2}{24a^4\lambda^2}\right)\right)\Big\},
\end{align}
which is clearly linear in the lapse function.

The conclusion of this section is that BIMOND suffers from ghost pathologies,  except in its constrained version, where it recovers the $f(\QGR)$ theory in the coincident gauge. Although, in this case it ceases to be a bimetric theory.

\section{Conclusion}
In this work we have considered a relativistic, ghost-free, and covariant hybrid-formulation of MOND in terms of a non-linear extension of the non-metricity scalar $f(\QGR)$. In the non-relativistic limit pertaining to small-velocity sources and weak field configurations one successfully recovers the MOND phenomenology on galactic scales. Moreover, a $\Lambda$CDM-like cosmological phenomenology can be assured on large cosmological scales. We confirmed its relation to constrained BIMOND theories, where the second metric is forced to be Minkowski and the potential interactions just the standard volume element. However, this can then barely be called a bimetric theory and loses its connection to the physical properties of BIMOND. Apart from this exceptional case where the $f(\QGR)$ theory is recovered, all BIMOND theories in their original formulation suffer from the Boulware-Deser ghost. This is sourced in a kinetic coupling of the two metrics as well as ghostly potential interactions not satisfying the specific structure of the dRGT potentials. Hence, BIMOND is not an acceptable relativistic formulation of MOND. However, $f(\QGR)$, being free of these pathologies, represents a promising exceptional route for embedding MOND into a relativistic theory and recovering the right phenomenology on cosmological scales.


\section*{Acknowledgements}
We would like to thank Mordehai Milgrom for very useful discussion and his comments on an early version of the draft.
LH is supported by funding from the European Research Council (ERC) under the European Unions Horizon 2020 research and innovation programme grant agreement No 801781 and by the Swiss National Science Foundation grant 179740. 


\end{document}